\documentstyle[12pt]{article}
  
\catcode`@=11
\def\secteqno{\@addtoreset{equation}{section}%
\def\theequation{\thesection.\arabic{equation}}}
\catcode`@=12
\secteqno
\newcommand{\be}{\begin{equation}}
\newcommand{\ee}{\end{equation}}
\newcommand{\bea}{\begin{eqnarray}}
\newcommand{\eea}{\end{eqnarray}}
\newcommand{\bref}[1]{(\ref{#1})}
\newcommand{\nn}{\nonumber}

\newcommand{\C}[1]{{\cal #1}}
\newcommand{\mapright}[1]{\smash{[mathop{\hbox to 1cm{\rightarrowfill}}
\limits^{#1}}}

\newcommand{\slM}{/ {\hskip-0.27cm{M}}}

\newcommand{\slT}{/ {\hskip-0.27cm{T}}}
\newcommand{\lw}[1]{\smash{\lower2.ex\hbox{#1}}}
\begin{document}
\thispagestyle{empty}
\hfill June 5, 2009

\hfill KEK-TH-1314
\vskip 20mm

\begin{center}
{\Large\bf Coset for Hopf fibration and Squashing}
\vskip 6mm
\medskip
\vskip 10mm
{\large Machiko\ Hatsuda$^{\ast\dagger}$
~and~
Shinya\ Tomizawa$^\ast$ \\ }
\parskip .15in
{\it $^\ast$Theory Division,\ 
High Energy Accelerator Research Organization (KEK),\\
\ Tsukuba,\ Ibaraki,\ 305-0801, Japan} \\
{\it $^\dagger$Urawa University, Saitama \ 336-0974, Japan}\\
\vskip 10mm
{\small e-mail:\ 
{mhatsuda@post.kek.jp} and
{tomizawa@post.kek.jp}}
\medskip
\end{center}

\vskip 10mm
\begin{abstract}
We provide a simple derivation of metrics for
fundamental geometrical deformations 
such as Hopf fibration, squashing and {\bf Z}$_k$ quotient 
which play essential roles in recent studies on the AdS4/CFT3.
A general metric formula of Hopf fibrations  
for complex and quaternion cosets is presented.
Squashing is given by a similarity transformation 
which changes the metric preserving the isometric symmetry of the
projective space.
On the other hand {\bf Z}$_k$ quotient is given as
a lens space which changes the topology  
preserving the ``local" metric.
\end{abstract}

\setcounter{page}{1}
\parskip=7pt
\newpage
\section{Introduction}

In recent studies on AdS4/CFT3 \cite{Aharony:2008ug}
interesting subjects are explored by fundamental geometrical deformations such as Hopf fibration, 
squashing and {\bf Z}$_k$ quotient
 \cite{Nilsson:1984bj,Awada:1982pk,Duff:1983nu}.
The AdS$_4\times${\bf S}$^7$ is a maximal supersymmetric solution of the M-theory,
and a gauge theory on the 3-dimensional 
membrane world volume including $N$=8 supersymmetries is expected to be its CFT dual
\cite{Aharony:2008ug,Bagger:2006sk}.
For the $N$=6 superconformal symmetry, whose bosonic subgroup
 is SU(4)$\times$U(1), 
the Chern-Simon-matter theory with level $k$
is conjectured to be CFT dual of the M theory
on the  AdS$_4\times${\bf S}$^7$/{\bf Z}$_k$ 
\cite{Aharony:2008ug}.
Dimensional reduction by increasing $k$ 
reduces it  into the type IIA superstring theory
on the  AdS$_4\times${\bf CP}$^3$ 
\cite{Aldazabal:2007sn,Gomis:2008jt}.
The $N$=1 superconformal theory with the bosonic subgroup 
Sp(2)$\times$Sp(1), where Sp(2)
=SO(5) and 
Sp(1)
=SO(3), 
is conjectured to be dual of
the M theory on  AdS$_4\times\tilde{\bf S}^7$/{\bf Z}$_k$
\cite{Ooguri} where
$\tilde{\bf S}^7$ is squashed {\bf S}$^7$. 
There is also squashed {\bf CP}$^3$ solution \cite{Ahn} 
 and other squashing gravity duals in the AdS/CFT 
 including $N$=2, 3 superconformal cases have been examined
\cite{Ahn:2008ua}.

Squashed solutions and {\bf Z}$_k$ quotient solutions
were 
widely studied in 
the 11-dimensional supergravity theory
\cite{Awada:1982pk,Duff:1983nu, 
Castellani:1983tc}
and in the superstring theories
\cite{Duff:1998cr}.
Recently intensive studies on such deformations
 have been explored 
in the 5-dimensional supergravity theory 
related to black hole  solutions
\cite{Ishihara:2005dp,Ishihara:2006iv}.
They are also useful to explore 11-dimensional  supergravity solutions
from the point of view of analogy between 5-dimensional 
 and 11-dimensional supergravity theories
\cite{Mizoguchi:1998wv}.
Squashing deforms a kind of 5-dimensional asymptotically flat black hole solution
into an asymptotically locally flat 
Kaluza-Klein black hole solution with a compact extra dimension.
{\bf Z}$_k$ quotient of a compact direction
gives a Kaluza-Klein monopole solution
with a magnetic charge $k$ \cite{Gross:1983hb}.
Squashing preserves the global isometric symmetry
but changes the local metric,
while {\bf Z}$_k$ quotient changes the topological structure
but does not change the local metric.

The Hopf fibration 
 describes a 3-dimensional sphere in terms of 
 a 1-dimensional fiber over a 2-dimensional sphere
 base space
 whose metric is a 
projective space metric locally.
Its generalization to 
real, complex, quarternionic and octonionic projective spaces 
exist.
The corresponding cosets are known as listed \cite{steenrod} 
in  table \ref{table1}. 
\begin{table}[htbp]
 \caption{Hopf fibration and coset}
 \begin{center}
  \begin{tabular}{|c|ccc|ccc|}
  \hline
\lw{\rm field}&\multicolumn{3}{|c|}{\rm Hopf~fibration}
&\multicolumn{3}{|c|}
{\rm coset~~~~~~~~~}\\
&{\rm total}&
{\rm fiber}&{\rm base}&{\rm sphere}&&{\rm projective~space}\\
\hline &&&&&&\\
{\rm real}&{\bf S}$^{N}$&
$\stackrel{\pm 1~{\rm points}}{\longrightarrow}$
&
{\bf RP}$^{N}$&
$\displaystyle\frac{{\rm O}(N+1)}{{\rm O}(N)}$&$\to$&
$\displaystyle\frac{{\rm O}(N+1)}{{\rm O}(N)\times{\bf Z}_2}$\\
&&&&&&\\
{\rm complex}&${\bf S}^{2N+1}$
&$\stackrel{{\Large{\bf S}}^1}{\longrightarrow} $
&${\bf CP}^N$&
$\displaystyle\frac{{\rm U}(N+1)}{{\rm U}(N)}$&$\to$&
 $\displaystyle\frac{{\rm U}(N+1)}{{\rm U}(N)\times{\rm U}(1)}$\\
&&&&&&\\
{\rm quaternion}&{\bf S}$^{4N+3}$ &
$\stackrel{{\bf S}^3}{\longrightarrow} $
&${\bf HP}^N$&
$\displaystyle\frac{{\rm Sp}(N+1)}{{\rm Sp}(N)}$&$\to$&
$ \displaystyle\frac{{\rm Sp}(N+1)}{{\rm Sp}(N)\times{\rm Sp}(1)}$\\
&&&&&&\\
\hline
  \end{tabular}
 \end{center}\label{table1}
\end{table}
It is denoted that
Sp($n$) 
 is a group of $n\times n$ matrices of quaternion numbers.  
In this paper we present a simple description of these spaces
by following the projective lightcone limit procedure
proposed in  \cite{Hatsuda:2002wf}.
The projective lightcone limit brings the AdS$_5$ space
into the 4-dimensional projective lightcone space 
where CFT lives in.
It was generalized to a Hopf fibration,
a map from a (${2N+1}$)-dimensional sphere 
to a ${N}$-dimensional complex projective space, 
as a simple derivation of the Fubini-Study metric
 \cite{Hatsuda:2007it}.
There is a gauged sigma model derivation of
{\bf CP}$^N$ metric \cite{Hitchin:1986ea}
where constrained $N+1$ complex coordinates are 
used.
Starting from the {\bf S}$^{2N+1}$ metric, 
a gauge field is introduced for the local U(1)
and then it is eliminated by the equation of motion
to obtain the {\bf CP}$^{N}$ metric.
Instead we use a whole U($N+1$) matrix as coordinates
which includes U(1) field and U($N$) fields.
In this formulation
it is clear how coset, sigma model
 and geometry are related; 
for a coset U($N+1$)/U($N$)
 the kinetic term for U($N$) fields is 
absent
resulting {\bf S}$^{2N+1}$,
and for a coset U($N+1$)/U($N$)$\times$U(1)
the kinetic term for U(1) field in addition to U($N$) fields is 
absent
resulting {\bf CP}$^{N}$.

In this paper we extend this  formulation to 
  Hopf fibrations
listed in table \ref{table1}
and to deformations such as
squashing and {\bf Z}$_k$ quotient.
In next section a general formula of Hopf fibrations
is derived where coordinates are embedded in a group matrix.
Squashing is explained as a similarity transformation.
After discussing a real coset case,
a complex coset case is  presented in section 4.
{\bf Z}$_k$ quotient is presented as a lens space there.
In section 5 a quaternion coset case is presented.
Squashed {\bf S}$^7$ in our formulation 
is also presented which is consistent to the one 
obtained by Awada, Duff and Pope \cite{Awada:1982pk}.
  
\par

\section{General formula}

In this section we present a simple derivation of 
metrics for a sphere and a projective space.
A squashing is introduced as a 
similarity transformation.

We begin by a metric for a sphere
 whose corresponding coset G/H is in table \ref{table1}.
For a general treatment we consider  embedding 
these groups, G and H, into orthogonal group O($m+n+1$) and O($n+1$)
where $m$ and $n$ take specific numbers as listed in 
table \ref{table2}.
\begin{table}[htbp]
 \caption{Embedding into orthogonal group of coset for sphere}
 \begin{center}
  \begin{tabular}{|c||c|c|c|c||c|}
    \hline
\begin{tabular}{c}
 orthogonal~group\\
 ~embedded~
 G/H
 \end{tabular}
    &     field   &  $m$  & $n$   &condition    & G/H
    \\
    \hline
    &&&&&\\
      &     real & 1 & $N-1$   &  none  &
         $ {\frac{{\rm O}(N+1)}{{\rm O}(N)}}$  \\
     $ {\displaystyle\frac{{\rm O}(m+n+1)}{{\rm O}(n+1)}}$  & 
       complex & 2   & $2N-1$   &  $g^tKg=K$  & 
       $\frac{{\rm U}(N+1)}{{\rm U}(N)}$
   \\       
        &    quarternion &  4& $4N-1$   &  
        $g^tKg=K$,~
$g^t{L} g={L}$  &   
        $\frac{{\rm Sp}(N+1)}{{\rm Sp}(N)}$
  \\&&&&&\\
    \hline
  \end{tabular}\label{table2}
 \end{center}
\end{table}
There are conditions on an element of
U($n$) and Sp($n$) as in  table
  \ref{table2} 
with
\bea
g^tKg=K~~,~~
K=\left(
\begin{array}{cccc}
\epsilon&0&\ldots &0\\
0&\epsilon&\ldots &0\\
\vdots &\vdots &\ddots &\vdots \\
0&0&\cdots &\epsilon
\end{array}
\right)~,~
\epsilon=\left(
\begin{array}{cc}
0&1\\
-1&0
\end{array}
\right)\label{Kahler}
\eea
and 
\bea
g^t{L} g={L}~~,~~{L}=
\left(
\begin{array}{cccc}
\epsilon_4&0&\cdots&0\\
0&\epsilon_4&\cdots&0\\
\vdots&&\ddots&\vdots\\ 
0&0&0&\epsilon_4
\end{array}
\right)~,~
\epsilon_4=
\left(\begin{array}{cc}
0&{\bf 1}_2\\-{\bf 1}_2&0
\end{array}
\right)
~.\label{LLL}
\eea

For an element of the coset 
embedded in O($m+n+1$) matrix 
$z_A{}^B$ with 
$A,B=0,\cdots,m,\cdots,m+n$
it is convenient to denote 
$z_A{}^0=x_A$.
The orthonormal condition, $z^tz={\bf 1}$, leads to 
the metric for a ($m+n$)-dimensional sphere
\bea
&(z^tz)^{00}=
\displaystyle\sum_{A=0}^{m+n} (x_A)^2=1&\nn\\
&ds^2_{{\bf S}^{m+n}}=
\displaystyle\sum_{A=0}^{m+n} (dx_A)^2
=\displaystyle\sum_{A,B=0}^{m+n} \delta^{AB}(J_A{}^0)^t J_B{}^0
\label{metricsphere}
\eea
where 
$J_A{}^B=(z^{-1}dz)_A{}^B$ is the left invariant (LI) one form.
This is invariant under the global  O($m+n+1$) transformation of $x_A$,
namely  a round $(m+n)$-dimensional sphere.  

On the other hand  the same isometric symmetry of the sphere, G,
is realized by a projective space 
as listed in \bref{table1}.
An element of G $\ni z$ is partitioned into four blocks
\bea
z=\left(
\begin{array}{cc}
A&B\\C&D
\end{array}\right)~~~,\label{ABCD}
\eea
where $A,~B,~C,~D$ are $m{\times}m$, $m{\times}(n+1)$,
 $(n+1){\times}m$, $(n+1){\times}(n+1)$
matrices respectively.
Diagonal blocks $A$ and $D$ are subgroup H
of the coset for a projective space, while 
only $D$ is included in H for a sphere coset.
One of the two off-diagonal blocks, $C$, contains
projective coordinates.
In order to describe a projective space 
it is convenient to parametrize $z$ as \cite{Hatsuda:2002wf}
\bea
z=
\left(
\begin{array}{cc}
1&0\\X&1
\end{array}\right)
\left(
\begin{array}{cc}
u&0\\0&v
\end{array}\right)
\left(
\begin{array}{cc}
1&Y\\0&1
\end{array}\right)~~.\label{zXYuup}
\eea 
The parameters in \bref{ABCD} and \bref{zXYuup}
are related as $u=A$ and $X=CA^{-1}$
showing that $X$ is projective.
Under the  G transformation,  
block coordinates are transformed as
\bea
&
z\to z'=\left(
\begin{array}{cc}
{\rm a}&{\rm b}\\{\rm c}&{\rm d}
\end{array}\right)
z~~,~~
\left(
\begin{array}{cc}
{\rm a}&{\rm b}\\{\rm c}&{\rm d}
\end{array}\right)
\in~{\rm G}
& \label{trans} \\
&
~~
\left\{\begin{array}{ccl}
X'&= &({\rm c}+{\rm d}X)({\rm a}+{\rm b}X)^{-1}\\
u'&=& ({\rm a}+{\rm b}X)u\\
v'&=& \left\{
{\rm d}-({\rm c}+{\rm d}X)({\rm a}+{\rm b}X)^{-1}{\rm b}
\right\}v\\
Y'&=&Y+u^{-1}({\rm a}+{\rm b}X)^{-1}{\rm b}v~~.
\end{array}\right.
&\nn
\eea
The  $X$ parameter
 is  a projective coordinate representing  G 
 by a fractional linear transformation.
The LI current is calculated as
\bea
&&~J=z^{-1}dz=
\left(
\begin{array}{cc}
J_u&J_Y\\J_X&J_{v}
\end{array}\right)
\label{JXuvY}
\\
&&\left\{\begin{array}{rcl}
J_X&=&v{}^{-1}dX u\\
J_u&=&
u^{-1}du-Yv{}^{-1}dX u\\
J_{v}&=&v{}^{-1}dv+v{}^{-1}dX uY\\
J_Y&=&dY+u^{-1}du Y-Yv{}^{-1}dv-Yv{}^{-1}dX uY~~.
\end{array}\right.~\nn~~\eea
Each element is invariant under G transformation \bref{trans}.

A sphere metric \bref{metricsphere}
 is generalized to 
$ds^2=\|J_X\|^2+\|J_u\|^2$ with a suitable norm.
The contribution of $\|J_Y\|^2$ equals to $\|J_X\|^2$
for the orthogonal group so we just use simpler expression 
  $\|J_X\|^2$ only.
The norm $\|w\|^2$ is defined in such a way that it equals to 
the norm  for a corresponding complex/quaternion number, $|w|^2$.
Complex conjugate ``$\ast$" is taken care  by transpose ``$t$" when
a complex matrix is  embedded in an real O($n$) matrix
as shown in the following sections. 
For example, suppose that $w$ is  O($2$) $2\times 2$ matrix written as
$a{\bf 1}_2+b\epsilon$ with real numbers $a,b$. The norm is defined as
$\|w\|^2=\frac{1}{2}{\rm tr}~ w^t w=a^2+b^2$ which is equal to
$|w|^2=w^\ast w=a^2+b^2$ when it is recognized as a complex number, $w=a+ib$.
We denote ``tr $M^{AB}$" as $\displaystyle\sum_{A=0}^{m-1}M^{AA}$,
since  we only need this trace for $m\times m$ matrices,
$(J_X{}^tJ_X)^{AB}$ and $(J_u{}^tJ_u)^{AB}$, with 
$m=2$ or $4$ for complex or quaternion cases respectively.
Both $\|J_X\|^2$ and $\|J_u\|^2$ are invariant under H transformation;
\bea
&z\to z'= z\left(
\begin{array}{cc}{ 1}&{ 0}\\{ 0}&h
\end{array}
\right)~~,~~
\left(
\begin{array}{cc}{ 1}&{ 0}\\{ 0}&h
\end{array}
\right)\in {\rm H}&\\
&J'_u= J_u~,~J'_X= h^{-1}J_X &\nn
\eea
and $h^th=1$ is used in the norm $\|J_X\|^2$ .

The orthonormal condition of $z$ of \bref{zXYuup}
is given by;
\bea
z^tz=1~\Rightarrow~
\left\{
\begin{array}{ccl}
uu^t&=&\left[1+X^tX\right]^{-1}\\
vv{}^t&=&\left[1-X(1+X^tX)^{-1}X^t\right]^{-1}\\
Y&=&-u^t X^t v
\end{array}
\right.\label{SOcondition}~~~.
\eea 
Inserting this relation into \bref{JXuvY},
the square of  $J_X$
is given by
\bea
||J_X||^2 &=&\frac{1}{m}{\rm tr}(J_X){}^t J_{X}
=\frac{1}{m}\displaystyle\sum_{B=0}^{m-1}
 \displaystyle\sum_{A=m}^{m+n} (J_X{}_A{}^B){}^t J_{X}{}_A{}^B\nn\\
 &=&\frac{1}{m}{\rm tr}\left(1+X^tX\right)^{-1}
  \left[dX^t dX -dX^t X\left(1+X^tX\right)^{-1}
X^tdX
\right]\nn\\
&=&
\displaystyle\frac{\|dX\|^2}{1+\|X\|^2}
-\displaystyle\frac{
\| X^t dX\|^2 }
{(1+\|X\|^2)^{2}}
~=~ ds^2_{\rm Fubini-Study}
 \label{afterHF}~~~.
\eea
This is a Fubini-Study metric for
a projective space \cite{Hatsuda:2007it}.
From the orthonormal condition of the first line of 
\bref{SOcondition} $u$ can be parametrized as
\bea
u=\frac{1}{\sqrt{1+\|X\|^2}}U~~~,~~~U^tU={\bf 1}_m~~~\label{uU}
\eea
where {\bf 1}$_m$ is a $m$-dimensional unit matrix.
The square of  $J_u$
is given by
\bea
||J_u||^2&=&
\frac{1}{m}{\rm tr}J_u{}^tJ_u
=
\frac{1}{m}{\rm tr}\left[u^{-1}du+u^tX^tdX u\right]^t
\left[u^{-1}du+u^tX^tdX u\right]\nn\\
&=&
\Biggr\|dU U^{-1}+
\displaystyle\frac{X^t dX-dX^\ast X}{2(1+|X|^2)}
\Biggr\|^2~=~
ds^2_{\rm Hopf-fiber}
~~~
\eea
It turns out that this is a metric for the Hopf-fiber.

Now let us perform a similarity  transformation 
with a parameter $\lambda$ as
\bea
z\to z_\lambda=
\Lambda
z
\Lambda^{-1}~~~,~~~
\Lambda=
\left(
\begin{array}{cc}
\lambda{\bf 1}_{m}
&0\\
0&{\bf 1}_{n+1}
\end{array}
\right)
~~~.\label{GL1}
\eea
Although this similarity transformation does not change the algebra,
it scales a part of LI one forms as
\bea
J~\to~
\left(
\begin{array}{cc}
J_u&\lambda J_Y\\J_X/\lambda&J_{v}
\end{array}\right)~~~.\label{lambdaJ}
\eea
Taking into account the rescaling
$ds^2\to ds^2/\lambda^2$ for a normalization,
the metric for a sphere becomes
\bea
&&ds^2_{{\bf  S}^{m+n}}~=~ds^2_{\rm Fubini-Study}
+\lambda^2 ds^2_{\rm Hopf-fiber}
\label{lambdametric}
\\\nn\\
&&\left\{\begin{array}{ccl}
ds^2_{\rm Fubini-Study}
&=&
\displaystyle\frac{|dX|^2}{1+|X|^2}
-\displaystyle\frac{
| X^\ast dX|^2 }
{(1+|X|^2)^{2}}
\nn\\
ds^2_{\rm Hopf-fiber}
&=&
\Biggr|dU U^{-1}+
\displaystyle\frac{X^\ast dX-dX^\ast X}{2(1+|X|^2)}
\Biggr|^2\end{array}\right.
\eea
where $X$ and $U$ variables are  complex or quaternion numbers
and the norm $\|w\|^2$ is replaced by the usual norm $|w|^2$;
for a complex case $2N\times 2$ rectangular matrix $X$  
is recognized as a complex $N$ vector,
and for a quaternion case $4N\times 4$ rectangular matrix is recognized as a quaternion 
$N$ vector.
The parameter  $\lambda$ in \bref{lambdametric}
is ``squashing" parameter which is in general a function of coordinates
\cite{Ishihara:2005dp,Ishihara:2006iv}.
For such cases the similarity transformation \bref{GL1} gives 
extra term in \bref{lambdaJ}, which
however is not added 
for preserving the G symmetry.
This transformation is different from the Weyl transformation 
so the Weyl tensor becomes non-vanishing in many cases 
which causes supersymmetry breaking.
The metric \bref{lambdametric}
 has G invariance independently on the value of $\lambda$
since each term is invariant under  G transformations \bref{trans}.
In addition to G symmetry
$||J_u||^2$ term has another symmetry 
which is a shift of  the Hopf fiber coordinate,
independently on the value of $\lambda$.

This deformation
 is recognized as the symmetry breaking of the orthonormal metric.
A round spehre metric \bref{metricsphere}
can be written in terms of the orthonormal frame as
\bea
ds^2_{{\rm round}~{\bf S}^{m+n}} 
&=&\frac{1}{m}\displaystyle\sum_{C=0}^{m-1}
\displaystyle\sum_{\underline{A},\underline{B}=0}^{m+n}\delta^{\underline{AB}}
(J_{\underline{A}}{}^C)^tJ_{\underline{B}}{}^C
\label{round}\nn\\
\delta^{\underline{AB}}&=&\left(
\begin{array}{cc}
{\bf 1}_{m}
&0\\
0&{\bf 1}_{n+1}
\end{array}
\right)~~~
\eea
where O($m+n+1$) symmetry is manifest. 
However the squashed sphere metric \bref{lambdametric}
gives the following expression
\bea
ds^2_{{\rm squashed}~{\bf S}^{m+n}}
&=&\lambda^2\|J_u\|^2+\|J_X\|^2\nn\\
&=&\frac{1}{m}\displaystyle\sum_{C=0}^{m-1}
\left(
\lambda^2 \displaystyle\sum_{{A},{B}=0}^{m-1}
\delta^{{AB}}(J_{{A}}{}^C{})^t
J_{{B}}{}^C{}+
\displaystyle\sum_{{A},{B}=m}^{m+n}
\delta^{{AB}}(J_{{A}}{}^C{})^t
J_{{B}}{}^C{}\right)
\nn\\
\delta^{\underline{AB}}&=&\left(
\begin{array}{cc}
\lambda^2{\bf 1}_{m}
&0\\
0&{\bf 1}_{n+1}
\end{array}
\right)~~~.\label{eta}
\eea
where  the symmetry  of the orthonormal metric is broken to
O($m$)$\times$O($n+1$).
Despite of this smaller vacuum symmetry
 the projective coordinates
can realize a larger symmetry G 
by the fractional linear transformation.  
Only for $\lambda^2=1$ the O($m+n+1$) invariance of the orthonormal metric 
 is recovered in addition to G symmetry.
In the $\lambda \to 0$ limit a sphere becomes a projective space with
the Fubini-Study metric \bref{afterHF}
where the space dimension is reduced.

\par

\section{Real coset}

At first we consider a real coset O($N+1$)/O($N$)
as {\bf S}$^0$ ($\pm 1$ points) fibration over  
{\bf RP}$^{N}$.
An element of G=O($N+1$)$\ni z$
is decomposed into four blocks as \bref{ABCD}.
In the parametrization of \bref{zXYuup}
 $u$ and $X$ are 
$1{\times}1$ and $N{\times}1$ matrices
respectively.
For example $N=7$ case, {\bf S}$^7$
is written as {\bf S}$^0$ ($\pm 1$) fibration over {\bf RP}$^7$.
A round seven sphere is easily seen in 
the parametrization
of $z\in$ O(8) as
\bea
z=
\left(
\begin{array}{c|c}
x_0&~~~~~~~~~~~~~~~~~~~~~~~~\\\hline
x_1&\\
x_2&\\
x_3&\\
x_4&\\
x_5&\\
x_6&\\
x_7&\\
\end{array}
\right)
\label{o8o7}
\eea
with $x_0=u$ and $x_A=X_A{}^0~u$ for $A=1,\cdots,7$,
while the Hopf fibration is easily seen in the {\bf RP}$^7$ 
coordinates $X_A{}^0$ with $A=1,\cdots,7$
and the fiber coordinate $u$.

The fiber coordinate ``$u$" in \bref{uU} can be chosen as
\bea
u=\frac{1}{\sqrt{1+X^2}}U~~,~~
X^2=\displaystyle\sum_{A=1}^nX_A{}^0X_A{}^0~~,~~U=\pm 1. \label{ureal}
\eea
The LI one form responsible for the fiber is zero, $J_u=0$. 
The metric \bref{lambdametric} becomes the one for 
{\bf RP}$^N$ is 
calculated  as
\bea
ds^2_{{\bf RP}^N}
&=&\frac{dX^2}{1+X^2}-\frac{(X\cdot dX)^2}{(1+X^2)^2}\nn\\
&=&\frac{dr^2}{(1+r^2)^2}+\frac{r^2 d\Omega_{(N-1)}^2}{1+r^2}\nn\\
&=&d\theta^2+\sin^2 \theta d\Omega^2_{(N-1)}
\eea
which is the metric for a ``locally" round ${\bf S}^N$.
In the second line from the bottom
$X^2=r^2$ and 
$dX^2=dr^2+r^2d\Omega_{(N-1)}^2$ are used and 
$r={\rm tan}\theta$ is used for the last line.
There is no room to introduce $\lambda$
for the {\bf S}$^0$ fiber.

\par

\section{Complex coset}

Next we consider a complex coset  U($N+1$)/U($N$) 
which corresponds to
{\bf S}$^1$ fibration over {\bf CP}$^{N}$.
 When an element of  U($N+1$) matrix is embedded in
  $(2N+2)$$\times$$(2N+2)$ orthogonal matrix, $z$, 
  it preserves
   a K\"{a}hler form metric 
 $z^t K z=K$ in  \bref{Kahler}
as well as $z^tz=1$ . 
Since $\epsilon^2=-1$ and $\epsilon^t=-\epsilon$,
$\epsilon$ is an imaginary base ``$i$"
and ``transpose" is replaced with complex conjugate ``$\ast$".
Then  $z$ is recognized as $(N+1){\times}(N+1)$ matrix of
complex numbers by recognizing a  
2$\times$2 matrix $w=a{\bf 1}_2+b\epsilon$ as a complex number $w=a+ib$
 with real numbers $a$ and $b$.

Let us decompose 
a coordinate $z\in$ U($N+1$)
into four blocks as
\bref{ABCD}.
$u$ and  $X$ in the parametrization \bref{zXYuup}
are $2\times 2$ and $2N\times 2$ matrices 
respectively.
{\bf S}$^7$ is given by $N$=3 which is  
{\bf S}$^1$ fibration over {\bf CP}$^3$.
A round seven sphere and
a  Hopf fibration 
are easily seen in the following
parametrizations of $z\in$ U(4)  respectively as
\bea
z=
\left(
\begin{array}{cc|ccc}
x_0&-x_1&~~~~~~&~~~~~~&~~~~~~\\
x_1&x_0&&&\\
\hline
x_2&-x_3&&&\\
x_3&x_2&&&\\
x_4&-x_5&&&\\
x_5&x_4&&&\\
x_6&-x_7&&&\\
x_7&x_6&&&
\end{array}\right)
\label{u4u3}
=\left(
\begin{array}{c|ccc}
~~~u~~~&~\begin{array}{c}~\\~\end{array}
~~~&~~~~~~&~~~~~~\\\hline
X_1~u&\begin{array}{c}~\\~\end{array}&&\\ 
X_2~u&\begin{array}{c}~\\~\end{array}&&\\ 
X_3~u&\begin{array}{c}~\\~\end{array}&&\\ 
\end{array}\right)
\label{uX3}~~~.
\eea

The fiber coordinate in \bref{uU} is parametrized as
\bea
u =
\displaystyle\frac{1}{\sqrt{1+\sum_{I=1}^{N}|X_I|^2}}
e^{i\phi}
~~~.\label{aphi}
\eea
From the metric formula
\bref{lambdametric}
the metric for a {\bf S}$^{2N+1}$ as 
{\bf S}$^1$ fibration over {\bf CP}$^{N}$ 
including a squashed parameter $\lambda$
is given as
\bea
&&ds^2_{{\bf S}^{2N+1}}~=~ds^2_{{\bf CP}^N}+ds^2_{{\bf S}^1}\label{S2N1}\\
&&\left\{\begin{array}{ccl}
ds^2_{{\bf CP}^N}&=&
\displaystyle\frac{\displaystyle\sum_{I=1}^N
\left|dX_I\right|^2}{1+\displaystyle\sum_{I=1}^N|X_I|^2}
-\displaystyle\frac{\Bigr|\displaystyle\sum_{I=1}^NdX_I{}^\ast X_I\Bigr|^2
}{\Bigl(1+\displaystyle\sum_{I=1}^N|X_I|^2\Bigl)^2}\nn\\
ds^2_{{\bf S}^1}&=&
\lambda^2 (d\phi +A)^2~~,~~
A=-
\displaystyle\frac{
\displaystyle
\sum_{I=1}^Ni\left(X^\ast_I dX_I-dX^\ast_I X_I\right)}{2\Bigl(1+\displaystyle
\sum_{I=1}^N\left|X_I\right|^2\Bigl)}
\end{array}\right.
~~~.\nn
\eea
$ds^2_{{\bf S}^{2N+1}}$ is invariant under  
the G=U($N$+1) symmetry 
with the transformation rule
 \bref{trans} independently from the value of $\lambda$,
where the {\bf CP}$^N$ coordinates are 
transformed as a linear fractional transformation.
In addition to  U($N+1$) there exists another U($1$) symmetry; 
$\phi\to \phi+c$ with constant number $c$.
Combining this U($1$) and U($1$) in U($N+1$),
the whole symmetry of the ($2N+1$)-dimensional sphere 
metric $ds^2_{{\bf S}^{2N+1}}$ in \bref{S2N1} is
SU($N+1$)$\times$U(1).

The Einstein condition allows only  $\lambda^2=1$  
which corresponds to a round {\bf S}$^{2N+1}$ metric
with O($2N+2$) symmetry. 
Reducing dimensions allows another Einstein solution with
 $\lambda=0$ corresponding to the Fubini-Study metric for
 {\bf CP}$^N$.
At $\lambda^2=1$ the  O(2$N$+2) symmetry is recovered 
from U($N$+1)
which is reflection of recovery of 
the vacuum symmetry as shown in \bref{eta}.
It is also easily seen by introducing the coordinates
$
\tilde{X}_I=X_I e^{i\phi}
$,
then the metric is rewritten as
\bea
ds^2_{{\bf S}^{2N+1}}&=&\frac{d\phi^2
+\displaystyle\sum_{I=1}^N
{\mid}d\tilde{X}_I{\mid}^2
}{1+\displaystyle\sum_{I=1}^N{\mid}\tilde{X}_I{\mid}^2}
-\frac{
\left(\frac{1}{2}d\displaystyle\sum_{I}^N
{\mid}\tilde{X}_I{\mid}^2
\right)^2
}{\left(1+\displaystyle\sum_{I=1}^N{\mid}\tilde{X}_I{\mid}^2\right)^2}
+(\lambda^2-1)
\displaystyle\frac{\left(d\phi +\tilde{a}\right)^2
}{\left(1+\displaystyle\sum_{I=1}^N{\mid}\tilde{X}_I{\mid}^2\right)^2}
~~~\label{lambda21}\\
\tilde{a}&=&-
\displaystyle\frac{1}{2}
\displaystyle\sum_{I=1}^Ni\left(\tilde{X}_I{}^\ast d\tilde{X}_I
-d\tilde{X}_I^\ast \tilde{X}_I
\right)\nn
\eea
At $\lambda^2=1$ with changing variables,
$\sum_I {\mid}\tilde{X}_I{\mid}^2=r^2$,
$\sum_I {\mid}d\tilde{X}_I{\mid}^2=dr^2+r^2d\Omega_{(2N-1)}$ and $r=\tan ^2\theta$,
the metric becomes
\bea
ds^2_{{\bf S}^{2N+1}}=d\theta^2+\sin^2 \theta d\Omega_{(2N-1)}^2
+\cos ^2\theta  d\phi^2 \label{SO8tilde}
\eea
describing a round {\bf S}$^{2N+1}$. 
The first two terms in 
\bref{lambda21} contain real symmetric combinations,
${\mid}V{\mid}^2$, whose symmetry
is not only U($N+1$) but enlarged to O($2N+2$).
On the other hand the third term in \bref{lambda21} contains
``imaginary" skew combinations $V^\ast W-W^\ast V$
whose invariance is Sp($2N+2$;{\bf R}). 
Requiring both invariances at $\lambda^2\neq 1$
reduces to U($N+1$) invariance.

Now let us consider {\bf Z}$_k$ quotient of a sphere
which is important deformation to change the topology
\cite{Aharony:2008ug, Ishihara:2006iv}.
Lens spaces, {\bf S}$^3$/{\bf Z}$_k$,
were considered for the geometrical interpretation of 
D-branes \cite{Maldacena:2001ky}.
We present a general description of lens spaces 
in our coset formalism.
Let $\omega$  be a primitive $k$-th root of unity, $\omega^k={\bf 1}$
and 
 $q_0,q_1,q_2,\cdots$ be coprime to $k$.
A lens space L($k;q_0,q_1,q_2,\cdots$) is a quotient of the $\omega$ action
for complex coordinates $Z_A$ as
$(Z_0,Z_1,Z_2,\cdots)$   
$\to$   
$(\omega^{q_0}Z_0,\omega^{q_1}Z_1,\omega^{q_2}Z_2,\cdots)$.
This is realized as the ``left" action in our formalism
as
\bea
z=\left(
\begin{array}{c|ccc}
Z_0&&&\\
Z_1&&&\\
Z_2&&&\\
\vdots&&&\\
\end{array}
\right)~\to~z_\omega &=&
\left(
\begin{array}{cccc}
\omega^{q_0}&&&\\
&\omega^{q_1}&&\\
&&\omega^{q_2}&\\
&&&\ddots\\
\end{array}
\right)
\left(
\begin{array}{c|ccc}
Z_0&&&\\
Z_1&&&\\
Z_2&&&\\
\vdots&&&\\
\end{array}
\right)~~~.
\eea 
Therefore it does not change the LI one form,
 $z_\omega{}^{-1}dz_\omega=z^{-1}dz$, and the ``local" metric.
But it changes periods of the fiber coordinates,
so it changes a topology in general.
A lens space L(2;1,$\cdots$,1)
is {\bf RP}$^{2N+1}$, and
lens spaces L($k$;$q_0,q_1,q_2,\cdots$)
are {\bf S}$^{2N+1}$/{\bf Z}$_k$.
We consider a lens space L($k$;1,$\cdots$,1) 
as {\bf S}$^{2N+1}$/{\bf Z}$_k$ \cite{Aharony:2008ug}.
Contrast to the squashing action \bref{GL1},
the {\bf Z}$_k$ quotient is obtained by the left action of 
$\omega=e^{i 2\pi/k}$ operation:
\bea
z=\left(
\begin{array}{c|ccc}
Z_0&&&\\
\hline
Z_1&&&\\
Z_2&&&\\
\vdots&&&\\
\end{array}
\right)~\to~z_\omega &=&
\left(
\begin{array}{c|ccc}
\omega&&&\\
\hline
&\omega&&\\
&&\omega&\\
&&&\ddots\\
\end{array}
\right)
\left(
\begin{array}{c|ccc}
Z_0&&&\\
\hline
Z_1&&&\\
Z_2&&&\\
\vdots&&&\\
\end{array}
\right)
\\
\nn\\
&=&
\left(
\begin{array}{c|ccc}
{\bf 1}&&{\bf 0}&\\
\hline
X_1&&&\\
X_2&&{\bf 1}&\\
\vdots&&& \\
\end{array}
\right)
\left(
\begin{array}{c|ccc}
\omega u&&{\bf 0}&\\
\hline
&&&\\
{\bf 0}&&{\bf 1}&\\
&&&\\
\end{array}
\right)\left(
\begin{array}{c|ccc}
{\bf 1}&&Y&\\
\hline
&&&\\
{\bf 0}&&{\bf 1}&\\
&&& \\
\end{array}
\right)\nn
\eea   
In the second line it is shown that
this action reduces into the action only on $u$. 
The  {\bf Z}$_k$ quotient $z\sim z_\omega$ 
reduces to  $u \sim \omega u$ 
which changes the boundary condition of $\phi$ introduced in \bref{aphi},
\bea
&u\sim \omega u \Leftrightarrow
\phi\sim \phi+\displaystyle\frac{2\pi}{k}
&~~~.
\eea
In general boundary conditions in lense space 
produce more general topological deformations. 

We can examine the effect of the {\bf Z}$_k$ quotient 
at $\lambda^2=1$ for example,
since the {\bf Z}$_k$ quotient is independent procedure from squashing.
For a round sphere metric \bref{SO8tilde} 
there is $d\phi^2$ term where
$\phi$ has period $2\pi/k$.   
If we rewrite it in terms of a $2\pi$ period coordinate $\varphi$,
then it becomes 
$d\phi^2~\to$ $d\varphi^2/k^2$ 
so the orthonormal metric 
is distorted as $\delta^{\underline{\phi}\underline{\phi}}=1$ $\to
\delta^{\underline{\varphi}\underline{\varphi}}=1/k^2$.
In the O($2+2N$) matrix coordinate
$d\phi$ is included only in the $ds^2_{{\bf S}^1}$, and 
the symmetry of the vacuum orthonormal metric is broken
to O($2$)$\times$O($2N$). 
It still preserve G=U($N+1$) symmetry manifestly independent from the 
value of $k$.
The metric for a {\bf S}$^{2N+1}$/{\bf Z}$_k$, which is 
{\bf S}$^1$/{\bf Z}$_k$ fibration over {\bf CP}$^{N}$, 
is given as
\bea
&&ds^2_{{\bf S}^{2N+1}/{\bf Z}_k}~=~ds^2_{{\bf CP}^N}+ds^2_{{\bf S}^1/{\bf Z}_k}
\label{S2N1Zk}\\
&&\left\{\begin{array}{ccl}
ds^2_{{\bf CP}^N}&=&
\displaystyle\frac{\displaystyle\sum_{I=1}^N
\left|dX_I\right|^2}{1+\displaystyle\sum_{I=1}^N|X_I|^2}
-
\frac{\Bigr|\displaystyle\sum_{I=1}^NdX_I{}^\ast X_I\Bigr|^2
}{\Bigl(1+\displaystyle\sum_{I=1}^N|X_I|^2\Bigl)^2}\\
ds^2_{{\bf S}^1/{\bf Z}_k}&=&
 (d\phi +A)^2~=~\displaystyle\frac{1}{k^2}\left(d\varphi +kA\right)^2\\\\
 A&=&-
\displaystyle\frac{
\sum_{I=1}^Ni\left(X^\ast_I dX_I-dX^\ast_I X_I\right)}{2(1+
\sum_{I=1}^N\left|X_I\right|^2)}
\end{array}\right.
~~~\nn
\eea
where $ds^2_{{\bf CP}^N}$ and $A$ are the same as the ones in  \bref{S2N1}.
This is locally a round metric so 
the same Ricci curvature and the same energy independently on the value of $k$.
But the isometry of this space is not O($2N+2$) but
breaks down to SU($N+1$)$\times$U(1) at $k\neq 1$.
Increasing $k$, the fiber contribution becomes small
and only {\bf CP}$^N$ space remains which satisfies the Einstein condition
in $2N$-dimensional space.  
\par

\section{Quaternion coset}

In the last we consider a quaternion coset Sp($N+1$)/Sp($N$)
which corresponds to {\bf S}$^3$ fibration over {\bf HP}$^N$.
When an element of Sp($N+1$) matrix is embedded in $(4N+4){\times}(4N+4)$ orthogonal matrix
$z$, it satisfies U($2N+2$) condition $z^tKz=K$ with \bref{Kahler}
 and the symplectic condition
$z^t{L} z={L}$ with \bref{LLL}.
Such matrix, $z$, is given by a $(N+1)\times(N+1)$ matrix 
where each element is $4\times 4$ matrix
$w=a{\bf 1}_4+b\hat{\tau}_1+
c\hat{\tau}_2+
d\hat{\tau}_3$ with real numbers $a,b,c,d$ and basis
\bea
\hat{\tau}_i=\left\{
\begin{array}{l}
\hat{\tau}_1=1\otimes \epsilon\\
\hat{\tau}_2=\epsilon\otimes \tau_1\\
\hat{\tau}_3=\epsilon\otimes \tau_3\\
\end{array}
\right.~~~.&
\eea
Since 
$\hat{\tau}_i\hat{\tau}_j=-\delta_{ij}+\epsilon_{ijk}\hat{\tau}_k$,
  $\hat{\tau}_1\hat{\tau}_2\hat{\tau}_3=-1$ and
$\hat{\tau}_i^t=-\hat{\tau}_i$,
$\hat{\tau_i}$ are quaternionic imaginary basis
and transpose is complex conjugate  ``$\ast$".
So $w$ is  a quaternion number giving 
 a real norm, 
 $\|w \|^2= \frac{1}{4}{\rm tr}~ w^tw=a^2+b^2+c^2+d^2$ 
 which is equal to $|w|^2=w^\ast w$.

Let us decompose
a coset element $z\in$ Sp($N+1$)
into four blocks as \bref{ABCD}.
$u$ and $X$ in the parametrization \bref{zXYuup}
are $4\times 4$ and $4N\times 4$ matrices 
respectively.
{\bf S}$^7$ is given by  $N=1$ case, which is 
{\bf S}$^3$ fibration over {\bf HP}$^1$.
An element $z\in$ Sp(2) is parametrized 
as follows:
\bea
z=\left(
\begin{array}{cccc|c}
x_0&-x_1&-x_2&-x_3&~~~~~~~~~~~~~~~~~~~\\
x_1&x_0&-x_3&x_2&\\
x_2&x_3&x_0&-x_1&\\
x_3&-x_2&x_1&x_0&\\
\hline
x_4&-x_5&x_6&x_7&\\
x_5&x_4&x_7&-x_6&\\
x_6&x_7&x_4&-x_5&\\
x_7&-x_6&x_5&x_4&\\
\end{array}\right)
\label{xembedded}\label{sp2sp1}
=
\left(
\begin{array}{c|c}
~~u~~& \begin{array}{c}~\\~\\~\\~\end{array}~~~~~~~~~~~~~~~~~~\\
\hline
~~Xu~~& \begin{array}{c}~\\~\\~\\~\end{array}\\
\end{array}\right)
\eea
The first parametrization is convenient 
for a round seven sphere and the second parametrization is
convenient for the Hopf fibration.

From the orthonormal condition 
u can be parametrized as \bref{uU}  
\bea
u=\displaystyle\frac{1}{\sqrt{1+\sum_{I=1}^N |X_I|^2}}U~~,~~
U\in {\rm Sp}(1)~~~.
\eea
From the metric formula \bref{lambdametric}
the  metric for a {\bf S}$^{4N+3}$ 
which is a {\bf  S}$^3$ fibration over {\bf HP}$^N$
is obtained as 
\bea
&&ds^2_{{\bf S}^{4N+3}}~=~
ds^2_{{\bf HP}^{N}}
+ds^2_{{\bf S}^{3}}\label{S3HPN}\\&&
\left\{\begin{array}{ccl}
ds^2_{{\bf HP}^{N}}&=&\displaystyle
\frac{\displaystyle
\sum_{I=1}^N
|dX_I|^2}{1+\displaystyle
\sum_{I=1}^N |X_I|^2}
-\displaystyle
\frac{\left|\displaystyle
\sum_{I=1}^N
dX_I{}^\ast X_I\right|^2}{(1+\displaystyle
\sum_{I=1}^N|X_I|^2)^2}
\nn\\\\
ds^2_{{\bf S}^{3}}&=&
\lambda^2 \displaystyle\sum_{i=1,2,3}(\nu_i+A_i)^2~~,~~
\nu_i~=~  \Re \left(dUU^{-1}~\hat{\tau}_i{}^{-1}
\right)\nn\\\\
A_i&=&\Re \left(\displaystyle\frac{\displaystyle
\sum_{J=1}^N
(X_J{}^\ast dX_J-dX_J{}^\ast X_J)}
{2(1+\displaystyle
\sum_{I=1}^N|X_I|^2)} ~\hat{\tau}_i{}^{-1}
\right)~~~
\end{array}\right.
 \eea
where $\Re(w)$ denotes a real part of $w$. 
This metric has  Sp($N+1$)$\times$Sp(1) symmetry 
independently from the value of $\lambda$;
The {\bf HP}$^N$ coordinates, $X_I$, 
are transformed as a linear fractional transformation given by \bref{trans}
realizing Sp($N+1$) symmetry.
The $ds^2_{{\bf S}^3}$ has additional Sp(1) symmetry
under which $X_I$'s are inert.

As in the general argument \bref{round}
it becomes a round ($4N+3$)-dimensional sphere 
at $\lambda^2=1$
and the global symmetry is enhanced to O($4N+4$).
At $\lambda^2=1$ 
the metric \bref{S3HPN} becomes the one for a
 round {\bf S}$^{4N+3}$ which is easily seen in
the following coordinates
$\tilde{X}_I=X_IU$ as
\bea
ds^2_{{\bf S}^{4N+3}}&=&
\frac{\displaystyle\sum_{i=1,2,3
}\nu_i{}^2+
\displaystyle\sum_{I}^N
|d\tilde{X}_I|^2}{1+\displaystyle\sum_{I=1}^N|\tilde{X}_I|^2}
-
\frac{\Bigl(\frac{1}{2}d\displaystyle\sum_{I=1}^N
|\tilde{X}_I|^2 \Bigl)^2}
{\Bigl(1+\displaystyle\sum_{I=1}^N|\tilde{X}_I|^2\Bigl)^2}\nn
\\
&=&d\theta{}^2+\sin^2 \theta d\Omega_{(4N-1)}^2
+\cos^2 \theta d\Omega_{(3)}^2
\label{313}
\eea 
where $\displaystyle\sum_{I=1}^N|\tilde{X}_I|^2=(\tan \theta)^2$ and 
$\displaystyle\sum_{I=1}^N|d\tilde{X}_I|^2/(1+|\tilde{X}|^2)=d\theta^2
+\sin\theta^2d\Omega_{(4N-1)}^2$ are used.

Next let us focus on $N=1$ case to examine a non-trivial 
squashed {\bf S}$^7$ solution.
In order to correspond a round {\bf S}$^7$ coordinates and 
a squashed {\bf S}$^7$ coordinates, 
 Sp(1)$\ni U$ and {\bf HP}$^1$  coordinate $X$ are expressed 
in terms of $x_A$, ${A=1,\cdots,7}$ with
$|X|^2=x_4{}^2$ and $X =x_4 V$
as
\bea
U=\displaystyle\frac{1+\displaystyle\sum_{i=1,2,3}x_i\hat{\tau}_i}{\sqrt{1+\displaystyle\sum_{l=1,2,3}x_l{}^2}}
~~,~~
V=\displaystyle\frac{1+\displaystyle\sum_{i=1,2,3}x_{4+i}\hat{\tau}_i}{\sqrt{1+\displaystyle\sum_{l=1,2,3}x_{4+l}{}^2}}
~~~.\nn
\eea
This is an embedding of an instanton solution on {\bf S}$^4$ into 
SU(2) fiber.
The metric for {\bf S}$^7$ as a {\bf S}$^3$ fibration over 
{\bf HP}$^1$ is given by
\bea
&&ds^2_{{\bf S}^7}~=~\frac{1}{4}d\theta{}^2
+\frac{\sin \theta{}^2}{4}\tilde{\nu}_i{}^2
+\lambda^2\displaystyle\sum_{l=1,2,3}
(\nu_i+A_i)^2\label{S7lambda}\\
&&\left\{\begin{array}{ccl}
A_i&=&\displaystyle
\frac{1-\cos \theta}{2}\tilde{\nu}_i
\\
\nu_i&=&\Re \left(dU U^{-1}~\hat{\tau}_i{}^{-1}
\right)=
\displaystyle\frac{1}{1+\displaystyle\sum_{l=1,2,3}x_l{}^2}
(dx_i+\epsilon_{ijk}x_jdx_k)\nn\\
\tilde{\nu}_i&=& \Re \left( V^{-1}dV~\hat{\tau}_i{}^{-1} 
\right)
=
\displaystyle\frac{1}{1+\displaystyle\sum_{l=1,2,3}x_{4+l}{}^2}
(dx_{4+i}-\epsilon_{ijk}x_{4+j}dx_{4+k})\end{array}\right.\nn
\eea
where $|X|^2=\left(\tan \frac{\theta}{2}\right)^2$
which is different from $\theta$ in \bref{313}. 
Two Sp(1) currents, 
 $\nu_i$ and $\tilde{\nu}_i$, satisfy
$\sum \nu_i{}^2=\sum \tilde{\nu}_i{}^2=
d\Omega_{(3)}{}^2$,
$d\nu_i=\epsilon_{ijk}\nu_j\wedge \nu_k$ and
$d\tilde{\nu}_i=-\epsilon_{ijk}\tilde{\nu}_j\wedge \tilde{\nu}_k$.
The metric in vielbein form is given as
\bea
ds^2_{{\bf S}^7}&=&e^{\underline{a}}\delta_{\underline{ab}}e^{\underline{b}}~~,~~\underline{a}=(\underline{1},\underline{2},\underline{3},\underline{\theta},\underline{5},\underline{6},\underline{7})\nn\\
e^{\underline{i}}&=&\lambda(\nu_i+\frac{1-\cos \theta}{2}\tilde{\nu}_i)\nn\\
e^{\underline{\theta}}&=&\frac{1}{2}d\theta\nn\\
e^{\underline{4+i}}&=&\frac{\sin \theta}{2}\tilde{\nu}_i~~~.
\eea
The Ricci tensor with local Lorentz indices are
\bea
R_{\underline{i}}{}^{\underline{j}}=(\frac{2}{\lambda^2}+4\lambda^2)\delta_{\underline{i}}^{\underline{j}}~~,~~R_{\underline{4+i}}{}^{\underline{4+j}}=6(2-\lambda^2)\delta_{\underline{i}}^{\underline{j}}~~,~~
R_{\underline{\theta}}{}^{\underline{\theta}}=6(2-\lambda^2)~~~.
\eea
The Einstein metric condition, $R_{\underline{ab}}=c \delta_{\underline{ab}}$ $\Leftrightarrow$
$R_{mn}=c g_{mn}$, equates these coefficients to be equal.
There are two solutions 
\bea
6(2-\lambda^2)=\frac{2}{\lambda^2}+4\lambda^2~~\to~~
\lambda^2=1~,~\displaystyle\frac{1}{5}~~~.
\eea
Detail of the computation is in the appendix.
The $\lambda^2=1$ solution corresponds to an O(8) invariant round seven sphere
solution,
while the $\lambda^2=1/5$ solution 
corresponds to a squashed seven sphere which 
has only Sp(2)$\times$Sp(1)$=$ SO(5)$\times$SO(3) invariance \cite{Awada:1982pk,Duff:1983nu}.
Our coordinate system in \bref{S7lambda} is slightly different from the one of 
the Awada, Duff and Pope, but
it gives the same result which is the same 
ratio of sizes of the fiber {\bf S}$^3$
and the base {\bf HP}$^1$={\bf S}$^4$.
For $\lambda^2=0$ 
it  becomes the Fubini-Study metric of {\bf HP}$^1$.
\par

\section{Conclusion and discussions}

We have presented a simple derivation of 
metrics for Hopf fibrations by a coset formulation. 
The coordinate is a group matrix
and it is decomposed into four blocks.
Fiber coordinates and projective coordinates are
treated differently by embedding into 
different blocks;
The former and the latter are embedded in
an upper-left diagonal block
and a lower-left off-diagonal block respectively.
The remaining diagonal block corresponds to the stability group.
Projective coordinates realize the isometric symmetry manifestly
by a fractional linear transformation.
Squashing is introduced as a similarity transformation
which preserves the isometric symmetry of the projective space.
It changes the metric and curvature tensors.
Squashed {\bf S}$^7$ is also obtained in this formulation 
which is consistent with the one \cite{Awada:1982pk}.
The {\bf Z}$_k$ quotient is introduced as a lens space
which also preserves the isometric symmetry of the projective space.
It does not change the ``local" metric 
but  changes the topology.
It may be interesting to examine general lens space solutions 
 {\bf S}$^{2N+1}$/{\bf Z}$_k$=L($k;q_0,q_1,\cdots,q_N$) where 
 $q_0,q_1,\cdots,q_N$ are coprime to $k$. 

Key of our simple description is the four blocks partition
of a coordinate matrix \bref{ABCD} and \bref{zXYuup} where 
the Hopf fiber coordinate
is embedded in the upper-left
diagonal block.
This is not the case 
for the supersymmetrization of 
AdS$_4\times${\bf S}$^7$ 
which is described by the supergroup OSp(8$\mid$4).
The  Hopf fibration breaks OSp(8$\mid$4) into
OSp(6$\mid$4) where OSp(6$\mid$4) is
 embedded in a diagonal block,
then the Hop fiber U(1) is not embedded in a diagonal block
 of this SO(8) spinor representation.
In the supergroup matrix there is another
U(1) under which OSp(6$\mid$4)  is singlet,  
so this is embedded in the diagonal block.
  There are several important U(1)'s in OSp(8$\mid$4)
clarified by  Gomis, Sorokin and Wulff
\cite{Gomis:2008jt}.
The Hopf fiber 
U(1) and  
U(1) in SU(4) which is SU(3) invariant are given by
\bea
T_{\rm Hopf-fiber}&=&M_{12}\nn\\
T_{\rm SU(3)-inv}&=&-3M_{12}+M_{34}+M_{56}+M_{78}~\Rightarrow~T_1\label{su3inv}
\eea
where $M_{ab}$ $a=1,\cdots,8$ are SO(8) generators
in the vector representation.
On the other hand the spinor representation is obtained by
multiplying gamma matrices $\Gamma_a$.
Spinor states are classified by the chirality operator
$\Gamma_1\cdots\Gamma_8$
and U(1) charge, $T_2$, 
\bea
T_2&=&
\slM_{12}+\slM_{34}+\slM_{56}+\slM_{78}\\
&\sim&
\sigma_3\otimes {\bf 1}\otimes {\bf 1}\otimes {\bf 1}
+{\bf 1}\otimes\sigma_3\otimes  {\bf 1}\otimes {\bf 1}+
 {\bf 1}\otimes {\bf 1}\otimes\sigma_3\otimes {\bf 1}+
{\bf 1}\otimes {\bf 1}\otimes {\bf 1}\otimes\sigma_3\nn\\
T_1&=&\slT_{\rm SU(3)-inv}=-3\slM_{12}+\slM_{34}+\slM_{56}+\slM_{78}~~~\nn
\eea
with $\slM_{12}=M_{12}\Gamma_{12}$. 
The U(1) generated by $T_2$ is embedded in a diagonal block.
$T_2$ corresponds to $K_{ab}\Gamma_{ab}$ in
\bref{Kahler}.
The positive chirality states are
$\mid \uparrow \uparrow \uparrow \uparrow\rangle$,
$\mid \downarrow\downarrow\downarrow\downarrow\rangle$,
$\mid \uparrow \uparrow \downarrow\downarrow\rangle $,
$\cdots$, $\mid \downarrow\downarrow\uparrow \uparrow\rangle $
with $T_2$ charge (2,$-2$,$0,0,0,0,0,0$) respectively, 
which is $1_2+1_{-2}+6_0$.
The negative chirality states are
$\mid \uparrow \uparrow \uparrow \downarrow\rangle$,$\cdots$,
$\mid \downarrow\downarrow\downarrow\uparrow\rangle$,$\cdots$
with $T_2$ charge ($1,1,1,1,-1,-1,-1,-1$) respectively,
which is
$4_1+4_{-1}$.
Survived $T_2$ invariant supergroup is OSp(6$\mid$4). 
The remaining spinor states with $T_2=\pm 2$ in the positive chirality sector
must make a closed supergroup OSp(2$\mid$4)
where this O(2)=U(1) generated by $T_2$.
Therefore the Hopf-fiber U(1) 
and another U(1) are given by
\bea
T_{\rm Hopf-fiber}&=&\frac{1}{2}(T_2-T_1)~=~
\slM_{12}
\\
T'&=&\frac{1}{2}(T_1+T_2)~=~
-\slM_{12}+\slM_{34}+\slM_{56}+\slM_{78}~~~.
\nn
\eea 
In the spinor representation $T_{\rm Hopf-fiber}$ can not be 
embedded in a block among four, because $|\slM_{12}^2|={\bf 1}_8$.
The bosonic part of  the coset is 
SU(4)$\times$U(1)/SU(3)$\times$U(1)',
where  U(1) in the numerator
is generated by $T_2$
and U(1)' in the denominator is generated by $T'$.
It will be useful to have a simple treatment of the supersymmetric 
Hopf fibration and deformations.
We put this problem for a future problem
including integrability  
analysis of the AdS4/CFT3 correspondence.

\section*{Acknowledgments}

We would like to acknowledge Shun'ya Mizoguchi and  Yosuke
Sumitomo for very useful  discussions and suggestions throughout this work.
M.H. is supported by the Grant-in-Aid for Scientific Research
No. 18540287. 
S.T. is supported by the JSPS under Contract No. 20-10616.

\vskip 10mm
\appendix

\section{Ricci tensor of squashed {\bf S}$^7$} 

The metric \bref{S7lambda} is written as
\bea
ds^2_{{\bf S}^7}&=&\frac{1}{4}d\theta{}^2
+\frac{\sin \theta{}^2}{4}\tilde{\nu}_i{}^2
+\lambda^2\displaystyle\sum_{l=1,2,3}
(\nu_i+\frac{1-\cos \theta}{2}\tilde{\nu}_i)^2~~~.\label{metsp4sp2}
\eea
In the vielbein is determined as 
$g_{mn}=e_m{}^{\underline{a}}e_n{}^{\underline{b}}\delta_{\underline{ab}}$
\bea
e_m{}^{\underline{a}}=
\left(
\begin{array}{ccc}
\lambda \Upsilon_i{}^{\underline{j}}&0&0\\
0&\displaystyle\frac{1}{2}&0\\
\hat{\lambda}\hat{\Upsilon}_{\hat{i}}{}^{\underline{{j}}}
&0&
\hat{\hat{\lambda}}\hat{\Upsilon}_{\hat{i}}{}^{\underline{\hat{j}}}
\end{array}
\right)~~,~~
\begin{array}{l}
m=(i,\theta,\hat{i})=(1,2,3,\theta,5,6,7)\\
\underline{a}
=(\underline{i},\underline{\theta},\underline{\hat{i}})=(\underline{1},\underline{2},\underline{3},\underline{\theta},\underline{5},\underline{6},\underline{7})
\end{array}
\eea
with
$
\hat{\lambda}=\lambda \displaystyle\frac{1-\cos \theta}{2}$
and 
$\hat{\hat{\lambda}}=\displaystyle\frac{\sin \theta}{2}
$
and
\bea
\begin{array}{cclcccl}
\Upsilon_{ij}&=&\displaystyle\frac{\delta_{ij}+\epsilon_{ijk}x^k}{1+\displaystyle\sum_{l=1,2,3}x_l{}^2}&,&
\hat{\Upsilon}_{ij}&=&\displaystyle\frac{\delta_{ij}-\epsilon_{ijk}x^{4+k}}{1+\displaystyle\sum_{l=1,2,3}x_{4+l}{}^2}
\\
\Upsilon^{-1}{}_{ij}&=&\delta_{ij}-\epsilon_{ijk}x^k+x_ix_j
&,&
\hat{\Upsilon}^{-1}{}_{ij}&=&\delta_{ij}+\epsilon_{ijk}x^{4+k}
+x_{4+i}x_{4+j}
\end{array}~~~.
\eea
The inverse vielbein is 
\bea
e_{\underline{a}}{}^m=
\left(
\begin{array}{ccc}
\displaystyle\frac{1}{\lambda} \Upsilon^{-1}{}_{\underline{i}}{}^{{j}}&0&0\\
0&2&0\\
-\displaystyle\frac{\hat{\lambda}}{\lambda\hat{\hat{\lambda}}}\delta_{\underline{\hat{i}}}^{\underline{j}}
{\Upsilon}^{-1}{}_{\underline{{j}}}{}^j
&0&
\displaystyle\frac{1}{\hat{\hat{\lambda}}}\hat{\Upsilon}_{\underline{\hat{i}}}{}^{\hat{j}}
\end{array}
\right)~~~.
\eea

Derivative operators in the local Lorentz frame indices are closed  as
\bea
e_{\underline{a}}=e_{\underline{a}}{}^m\partial_m~~,~~
\left[e_{\underline{a}},e_{\underline{b}}\right]=c_{\underline{a}\underline{b}}{}^{\underline{c}}e_{\underline{c}}~~,~~c_{\underline{a}\underline{b}}{}^{\underline{c}}=
-e_{\underline{a}}{}^me_{\underline{b}}{}^n\partial_{[m}e_{n]}{}^{\underline{c}}~~~.
\eea
Covariant derivative operators
\bea
\nabla_{\underline{a}}=e_{\underline{a}}{}^m\partial_m
+\frac{1}{2}\omega_{\underline{a}}{}^{\underline{b}\underline{c}}M_{\underline{c}\underline{b}}
\eea
satisfy
\bea
\left[\nabla_{\underline{a}},\nabla_{\underline{b}}\right]=T_{\underline{a}\underline{b}}{}^{\underline{c}}e_{\underline{c}}+\frac{1}{2}R_{\underline{a}\underline{b}}{}^{\underline{c}
\underline{d}}
M_{\underline{d}\underline{c}}
\eea
with Lorentz generator $M_{\underline{a}\underline{b}}$.
If the torsion is zero, $T_{\underline{a}\underline{b}}{}^{\underline{c}}=0$,
curvature is written in terms of the structure constant
\bea
\omega_{\underline{abc}}=\frac{1}{2}(c_{\underline{bca}}-c_{\underline{abc}}+c_{\underline{acb}})
~~,~~
R_{\underline{ab}}{}^{\underline{cd}}=
e_{[\underline{a}}\omega_{\underline{b}]}{}^{\underline{cd}}
-c_{\underline{ab}}{}^{\underline{e}}\omega_{\underline{e}}{}^{\underline{cd}}
+\omega_{[\underline{a}}{}^{\underline{ce}}\omega_{\underline{b}]\underline{e}}{}^{\underline{d}}~~~.
\eea

The covariant derivative operators for the Sp(2)$\times$Sp(1) space \bref{metsp4sp2},
which is torsionless, are computed as
\bea
\nabla_{\underline{i}}&=&\displaystyle\frac{1}{\lambda}\Upsilon^{-1}{}_{\underline{i}}{}^j
\partial_j+\frac{1}{2}\left\{
\displaystyle\frac{1}{\lambda}\epsilon_{\underline{ijk}}M_{\underline{jk}}
+\lambda\epsilon_{\underline{i\hat{j}\hat{k}}}
M_{\underline{\hat{j}\hat{k}}}-\lambda\delta_{\underline{i\hat{j}}}M_{\underline{\theta\hat{j}}}
\right\}\nn\\
\nabla_{\underline{\theta}}&=&2\partial_\theta\label{covder}\\
\nabla_{\underline{\hat{i}}}&=&
\displaystyle\frac{1-\cos\theta}{\sin \theta}\delta_{\underline{\hat{i}}}^{\underline{j}}
\Upsilon^{-1}{}_{\underline{j}}{}^k
\partial_k
+\displaystyle\frac{2}{\sin \theta}\hat{\Upsilon}^{-1}{}_{\underline{\hat{i}}}
{}^{\hat{j}}
\partial_{\hat{j}}\nn\\
&&
+\frac{1}{2}\left\{
\displaystyle\frac{2(1-\cos\theta)}{\sin\theta}\epsilon_{\underline{ijk}}M_{\underline{jk}}
-\lambda\epsilon_{\underline{\hat{i}\hat{j}{k}}}M_{\underline{\hat{j}{k}}}
-\displaystyle\frac{2}{\sin\theta}\epsilon_{\underline{\hat{i}\hat{j}\hat{k}}}M_{\underline{\hat{j}\hat{k}}}
-\lambda\delta_{\hat{i}k}M_{\underline{\theta k}}
-\displaystyle\frac{2\cos\theta}{\sin\theta}\delta_{\underline{\hat{i}\hat{k}}}M_{\underline{\theta \hat{i}}}
\right\}\nn~~~.
\eea
The Ricci tensor with local Lorentz indices, $R_{\underline{a}}{}^{\underline{b}}=
R_{\underline{ac}}{}^{\underline{bc}}$,  
\bea
&R_{\underline{i}}{}^{\underline{j}}=(\frac{2}{\lambda^2}+4\lambda^2)\delta_{\underline{i}}^{\underline{j}}~~,~~R_{\underline{4+i}}{}^{\underline{4+j}}=6(2-\lambda^2)\delta_{\underline{i}}^{\underline{j}}~~,~~
R_{\underline{\theta}}{}^{\underline{\theta}}=6(2-\lambda^2)&~~~\nn\\
&{\rm others}=0~~~.&
\eea
The Ricci tensor with curved indices is 
$R_{mn}=e_m{}^{\underline{a}}e_n{}^{\underline{b}}R_{\underline{ab}}$.


\end{document}